# On the Bandstructure Velocity and Ballistic Current of Ultra Narrow Silicon Nanowire Transistors as a Function of Cross Section Size, Orientation and Bias

Neophytos Neophytou*, Sung Geun Kim**, Gerhard Klimeck**, and Hans Kosina*

*Technical University of Vienna, TU Wien, Vienna, 1040, Austria.

**Network for Computational Nanotechnology, Birk Nanotechnology Center, Purdue University, West Lafayette, Indiana 47907-1285

*Corresponding author's email: neophytou@iue.tuwien.ac.at

## ABSTRACT

A 20 band $sp^3d^5s^*$ spin-orbit-coupled, semi-empirical, atomistic tight-binding (TB) model is used with a semi-classical, ballistic, field-effect-transistor (FET) model, to theoretically examine the bandstructure carrier velocity and ballistic current in silicon nanowire (NW) transistors. Infinitely long, uniform, cylindrical and rectangular NWs, of cross sectional diameters/sides ranging from 3nm to 12nm are considered. For a comprehensive analysis, n-type and p-type metal-oxide-semiconductor (NMOS and PMOS) NWs in [100], [110] and [111] transport orientations are examined. In general, physical cross section reduction increases velocities, either by lifting the heavy mass valleys, or significantly changing the curvature of the bands. The carrier velocities of PMOS [110] and [111] NWs are a strong function of diameter, with the narrower $D$=3nm wires having twice the velocities of the $D$=12nm NWs. The velocity in the rest of the NW categories shows only minor diameter dependence. This behavior is explained through features in the electronic structure of the silicon host material. The ballistic current, on the other hand, shows the least sensitivity with cross section in the cases where the velocity has large variations. Since the carrier velocity is a measure of the effective mass and reflects on the channel mobility, these results can provide insight into the design of NW devices with enhanced performance and performance tolerant to structure geometry variations. In the case of ballistic transport in high performance devices, the [110] NWs are the ones with both high NMOS and PMOS performance, as well as low on-current variations with cross section geometry variations.

I. Introduction

*Motivation:* The recent advancements in process and manufacturing of nanoelectronic devices have allowed the manufacturability of nanowire (NW) devices, which are considered as possible candidates for a variety of applications. For field-effect-transistor applications, nanowires have recently attracted large attention because of the possibility of enhanced electrostatic control, and the possibility of close to ballistic transport [1]. Ultra scaled nanowire transistors of diameters down to $D$=3nm and gate lengths down to $L_G$=15nm, have already been demonstrated by various experimental groups [2, 3, 4, 5, 6, 7]. Beyond the use in ultra-scaled high performance logic and memory transistors, NWs have also attracted large attention as biological sensors [8], optoelectronic [9, 10] and thermoelectric devices [11, 12]. Nanowire properties can be engineered and optimized through size, crystal orientation and strain [13, 14]. The carrier mobility and electrical conductivity that determine to a large degree the on-current and performance of devices, are quantities closely related to the bandstructure velocity of the channel. A thorough understanding of the bandstructure velocity and the parameters that control it, will aid the optimization and design of devices for a variety of electronic transport applications ranging from the diffusive to the ballistic limits. The bandstructure velocity and ballistic on-current in NWs as a function of the cross sectional size, transport orientation, carrier type, and gate bias, is the focus of this work.

The properties of 1D silicon NWs [13, 14, 15, 16, 17, 18] and 2D thin-body devices [19, 20, 21] in the high symmetry orientations [100], [110] and [111], have been addressed in a variety of theoretical studies. The challenges in simulating ultra-thin-body (UTB) and NW devices in which the atoms are countable in their cross section, call for sophisticated models beyond the effective mass approximation, especially in describing the valence band. The tight-binding [13, 14, 16, 17] and the k.p [20, 22] methods were used to calculate the electronic properties of these nanostructures, both with unstrained and strained lattice and scattering considerations [23, 24, 25, 26]. The performance in terms of on-current is associated with the carrier velocities, which are linked to the effective mass and the carrier mobility. Strain engineering, in that respect, is introduced in devices as a way to increase the carrier velocities and improve performance [27, 28].

These theoretical studies, however, have been performed for specific NW cross sections or thin-body widths. A comprehensive study that addresses the carrier velocities and ballistic on-current in NWs as a function of: i) channel cross sectional size, ii) carrier type (n-type or p-type), iii) transport orientation ([100], [110], [111]), iv) gate bias, and v) cross sectional shape (cylindrical/rectangular), has not yet been reported. Such a comprehensive device exploration will provide useful insight into optimization strategies of NW device performance for a variety of applications. The analysis presented in this paper addresses all these design factors. The nearest-neighbor atomistic tight binding (TB) model ($sp^3d^5s^*$-SO) [29, 30, 31, 32] is used for the NWs' electronic structure calculation, coupled to a 2D Poisson solver for the electrostatic potential. To evaluate transport characteristics, a simple semi-classical ballistic model [33, 34, 13, 14] is used. Cylindrical NWs of diameters from $D$=3nm to $D$=12nm, and rectangular NWs from 3nm to 12nm wide/tall (all combinations of aspect ratios) in three different transport orientations are considered. The design space could be further expanded by the use of strain as partly shown in Ref. [15], but this is beyond the scope of this paper.

We find that cross section reduction introduces changes in the bandstructure features that in general increase the carrier velocities. The increase can vary from ~0 to ~100% depending on the NW category. High inversion conditions (large gate biases) also increase the carrier velocities by ~50% as higher energy states are occupied. Device designs are identified for optimized performance, as well as performance variation tolerance to cross section variations. We note here that the ballistic model used in this study provides the upper limit for the performance of the devices. In reality, even devices with ultra short channel lengths are not 100% ballistic. The carrier velocity trends, however, reflect on the effective mass and carrier mobility and point to the direction of enhanced performance. In addition, imperfections and non-idealities will also affect the performance. In the last section of the paper, therefore, the results for one particular geometry are quantified in the presence of surface roughness scattering (SRS) using a quantum ballistic, full band, atomistic tight-binding simulator [17, 35, 36], and the magnitude of this additional variation is estimated.

This paper is organized as follows: In part II we describe the approach followed. In part III we present the results for the velocity and on-current as function of diameter in cylindrical

NWs. In part IV we discuss and provide explanations for the results. In part V we extend our analysis to rectangular NW devices. In part VI we provide design considerations for optimized performance. Finally, part VII summarizes and concludes the work.

## II. Approach

*Atomistic modeling:* To obtain the bandstructure of the NWs both for electrons and holes for which spin-orbit coupling is important, a well calibrated atomistic model is used. The nearest neighbor TB $sp^3d^5s^*$-SO model captures all the necessary band features, and in addition, is robust enough to computationally handle larger NW cross sections as compared to ab-initio methods. As an indication, the unit cells of the NWs considered in this study contain from ~150 to ~6500 atoms, and the computation time needed varies from a few hours to a few days for each bias point on a single CPU. The model itself and the parameterization used [29], have been extensively calibrated to various experimental data of various natures with excellent agreement [37, 38, 39, 40]. In particular, we highlight the match to experimental data considering the valley splitting in slanted strained Si ultra-thin-body devices on disordered SiGe [39], and single impurities in Si [40], without any material parameter adjustments. The model provides a simple but effective way for treatment of the surface truncation by hydrogen passivation of the dangling bonds on the surface of the nanowire [41]. What is important for this work, the Hamiltonian is built on the diamond lattice of silicon, and the effect of different orientations and cross sectional shapes is automatically included, all of which impact the interaction and mixing of various bulk bands.

*The simulation approach:* The devices simulated are cylindrical and rectangular NWs in the [100], [110] and [111] transport orientations, surrounded by $SiO_2$. In the case of the cylindrical devices, the oxide thickness is set to 8nm. These are typical NW device sizes that have been reported in experimental studies [3, 4]. In the rectangular cases, the oxide used is 1.1nm, which is a more realistic thickness for ultimately desired high performance devices. The simulation procedure consists of three steps as described in detail in Ref. [13] and is summarized here:

1. The bandstructure of the wire is calculated using the sp$^3$d$^5$s*-SO model. As an example, Fig. 1a and 1b show the conduction and valence bands of the $D$=6nm cylindrical NW in the [110] direction (positive-k parts of the dispersions). The different valleys, with different effective masses, as well as band interactions, non-parabolicities and anisotropies of the Si bulk bandstructure, especially for the valence band, are all captured by the model and appear in the NW dispersions.

2. A semi-classical top-of-the-barrier ballistic model is used to fill the electronic states and compute the transport characteristics [13, 14, 33, 34]. The range of validity of this model is explored in detail in Ref. [42] with a much more computationally demanding 3D full NEGF (Non-Equilibrium Green's Function) model. There it was shown that the approach is valid when the wire's length/width ratio exceeds $L/W$>5 in which case source-drain tunneling is suppressed.

3. The 2D Poisson equation is solved in the cross section of the wire to obtain the electrostatic potential. It is added to the diagonal on-site elements of the atomistic Hamiltonian as an effective potential for recalculating the bandstructure until self consistency is achieved.

Although the transport model used is a simple ballistic model, it allows for examining how the bandstructure alone will affect the channel properties and transport characteristics, and thus, it provides essential physical insight. It is the simplicity of the transport model, which allows to shed light on the importance of the dispersion details and their effect on transport.

### III. Results

The average carrier velocity (or injection velocity $v_{inj}$) of the wires, defined as $I_D/qn$, where $I_D$ is the ballistic drain on-current and $n$ is the carrier density in the NW cross section, depends strongly on carrier type, cross section diameter and orientation. Figure 2 presents these results, while Fig. 3 extends the analysis by including the dependence on gate bias as well. Figure 2a shows the carrier velocity for the cylindrical NMOS (dotted lines) and PMOS (circled lines) NWs in the [100] (blue), [110] (red) and [111] (black) transport orientations at on-state ($|V_G|=|V_D|=1$V) as a function of the wire's diameter. Large differences in the carrier velocities of the different NW orientation and carrier type are observed. Within the same NW category, large variations are also observed as the diameter reduces. The changes are more prominent in the PMOS [111], PMOS [110] and NMOS [110] cases. There, the velocity increases by up to a factor of two as the diameter reduces. The other three cases show minor velocity variations. As it will be explained in the "Discussion" section, this behavior originates purely from bandstructure changes.

Figure 2b shows the charge density as a function of the diameter. The charge increases almost linearly with cross section for devices at these diameter scales, following the increase in the oxide capacitance $C_{OX} = 2\pi\kappa\varepsilon_0/\ln\left(\frac{a+t_{OX}}{a}\right)$ with the radius $a$. The charge, however, is of similar magnitude in all cases, irrespective of orientation, NMOS or PMOS, because the gate oxide dominates the overall capacitance in Si NWs. (The NMOS NWs have only slightly higher charge than the PMOS ones). The total gate capacitance $C_G$ in our simulations is degraded from $C_{OX}$ by ~20-30% as also shown in Ref. [13]. This degradation is the same irrespective of NW type. NWs of different orientations or carrier type, have small differences in their quantum capacitance $C_Q$ of the order of <10% in most cases, but these cause only small differences in the total gate capacitance $C_G$ [13, 14, 15]. At $V_G=V_D=1$V, $C_Q$ varies from $C_Q$~1.5 nF/m to $C_Q$~9nF/m as the diameter increases from $D=3$nm to $D=12$nm. (It increases with gate bias because $C_Q$ a measure of the density of states at the Fermi level, and more and more subbands are occupied as the bias increases). The values are very similar in magnitude for all NW typs. The oxide capacitance, on the other hand, increases from $C_{OX} = 0.124$ nF/m to $C_{OX} = 0.26$ nF/m in the same diameter range, values 12 X and 34 X smaller than $C_Q$ respectively. This indicates that $C_{OX}$ still dominates the electrostatics (more for the larger diameters than the smaller ones). Of course the effect of $C_Q$ is more prominent for smaller oxide thicknesses. Still, the importance of $C_Q$ is reduced, and at a specific diameter, the already small variations between the different NW types, do not cause any variations in the charge density.

When considering ultrascaled transistors, transport can be close to the ballistic limit [4]. In this case the current through a device is simply given by the product of charge x velocity, $I_D =$ $qn \times v_{inj}$. The on-current vs. diameter is shown in Fig. 2c. It increases with diameter, but its magnitude, as well as its rate of increase is different for each NW case. The NWs with the larger velocity variations have the smaller variations in $I_{ON}$ as the diameter changes, as indicated by the PMOS [111] $\Delta I_D$ range vs. the NMOS [100] range in Fig. 2c. When the carrier velocity does not vary significantly with diameter, the change in $I_D$ follows the change in the charge density with diameter. This is the case for all NMOS NWs and the PMOS [100] NWs. In the PMOS [110] and [111] cases where the carrier velocity increases as the diameter decreases, the counter-acting effect of velocity increase and capacitance decrease makes the ballistic current more tolerant to diameter variations. This behavior is better illustrated in Fig. 2d, which shows the on-current of each NW category normalized to its higher value (the value of the $D$=12nm NW). The PMOS [110] and PMOS [111] NWs have the least on-current reduction as the diameter decreases. As the diameter is scaled by four times (from $D$=12nm to $D$=3nm), $I_D$ reduces by ~2 in the PMOS [110] and PMOS [111] cases, whereas it reduces by ~3 times in the rest of the NW categories. This behavior can potentially provide a mechanism for device designs with ballistic performance more tolerant to structural variations, especially considering the difficulties in controlling the line etch roughness in nanofabrication processes.

From Fig. 2a and 2c, it is evident that the designs are diameter dependent since at a certain NW diameter different NWs perform differently. For the NMOS cases, at larger diameters, the [100] NWs (dotted-blue) perform better, closely followed by the [110] NWs (dotted-red). At smaller diameters this order is reversed. For the PMOS cases, at larger diameters the performance of the [110] NW (circled-red) suffers compared to the [111] NWs (circled-black). At smaller diameters, however, the [110] NWs suffer a smaller performance reduction, and their performance becomes similar to the [111] NWs' performance. The fact that the NMOS [111], and especially the PMOS [100] NWs always lack on performance, leaves the [110] direction the one with high performance for both carrier types in the entire diameter range. This is enhanced by the fact that [110] NWs suffer less performance loss as the diameter reduces, for both carrier types. In CMOS applications, for which both NMOS and PMOS need to be utilized, [110] seems to be the optimal case.

The results of Fig. 2 are drawn at on-state at a fixed $|V_G|=|V_D|=1$V. Figure 3 generalizes these results by showing the variation of the velocity, charge and current as a function of diameter, orientation, and additionally gate bias. Row-wise, results for *carrier velocity* (first row), *charge* (second row), and *current* (third row) vs. $V_G$ are shown. Column-wise, results for the different transport orientations are shown: [100] – first column, [110] – second column, and [111] – third column. The left/right panels of each figure (negative/positive $V_G$), show results for PMOS/NMOS NWs respectively. The NWs considered are cylindrical in cross section (as shown in the inset of Fig. 3a) with diameters varying from $D$=12nm down to 3nm in decrements of 1nm. The arrows in each sub-figure indicate the direction of diameter reduction.

The carrier velocities of the NWs vs. $V_G$ are shown in Figures 3a, 3b, and 3c. In all cases, the carrier velocities increase with increasing $V_G$ (positive for the NMOS and negative for the PMOS cases). At higher inversion conditions, higher energy states are occupied with higher carrier velocity ($v_{inj} \sim dE/dk$). The increase in the carrier velocity with gate bias increase is as high as ~50%, and appears in all NW cases.

The strong diameter dependence in the cases of PMOS [110] and [111] NWs (Fig. 3b, 3c, left panels) is also observed through all biases. As the diameter scales from $D$=12nm to $D$=3nm, the velocity doubles from $v_{inj} \sim 0.7 \times 10^5$ m/s to $v_{inj} \sim 1.4 \times 10^5$ m/s (values around $V_G$=0V), independent of gate bias. A similar variation trend, but at a smaller scale is observed for the NMOS [110] NWs (Fig. 3b, right panel), for which the velocity increases from $v_{inj} \sim 0.9 \times 10^5$ m/s to $v_{inj} \sim 1.2 \times 10^5$ m/s as the diameter is reduced (values around $V_G$=0V). On the other hand, in the cases of PMOS [100] (Fig. 3a, left panel), NMOS [100] (Fig. 3a, right panel), and NMOS [111] (Fig. 3c, right panel), the carrier velocity has only a small diameter dependence. In these cases, the variation trends intermix at different gate biases, and cannot be identified as monotonically increasing, or decreasing with diameter. Explanations for all these trends are provided below in the "Discussion" section.

Comparing the magnitude of the velocities in each NW category, NMOS NWs in the [100] and [110] directions have similar performance, with their velocities at low gate bias being about $v_{inj} \sim 1 \times 10^5$ m/s. In the NMOS [110] case, the $D$=3nm NW has a slight advantage with the velocity raising to $v_{inj} \sim 1.15 \times 10^5$ m/s. The NMOS [111] NW velocities are ~20% lower at $v_{inj} \sim$ $0.8 \times 10^5$ m/s. In the PMOS NW cases, at a specific diameter size, the [111] NWs perform better, followed by the [110] NW and finally by the [100] oriented NWs. The PMOS [100] NWs can only deliver $v_{inj} \sim 0.5 \times 10^5$ m/s (low gate bias value), which makes them the NWs with the lowest carrier velocities of all categories examined for all diameter sizes. This behavior holds for all gate bias conditions.

In Fig. 3d-3f (second row) and Fig. 3g-3i (third row), we show the charge and ballistic current respectively, of the NWs with diameters from $D=8$nm down to $D=3$nm for which a large velocity variation is observed. As also shown in Fig. 2b, at the same gate overdrive the charge in the NWs of the same diameter is of similar magnitude, changing linearly with the oxide capacitance, irrespective of orientation, NMOS or PMOS.

The ballistic $I_D$ vs. $V_G$ characteristics in Fig. 3g, 3h and 3i are given by the product of the charge times velocity $I_D = qn \times v_{inj}$. In the NMOS NW cases (all right panels), where the carrier velocity does not vary significantly with diameter, the change in $I_D$ follows the change in the charge density with diameter. Same happens to the PMOS [100] NWs (Fig. 3g, left panel). At a given $V_G$ value, as the diameter decreases from $D=8$nm to $D=3$nm, the $I_D$ is almost halved. In the PMOS [110] and [111] cases Fig. 3h, 3i (left panels) where the carrier velocity increases as the diameter decreases, the ballistic current is tolerant to diameter variations and reduces only by ~20% and 30% respectively. This is a general behavior at all gate biases.

## IV. Discussion

The velocity variation trends with diameter and orientation are explained in Fig. 4 for NMOS and Fig. 5 for PMOS NWs. These figures show the first occupied subband (subband envelopes) of the NWs of each diameter at off-state conditions. Figures 4a, 4b and 4c show results for NMOS NWs in [100], [110], and [111] orientations, respectively. For example, Fig. 4a shows the first occupied subband of the $D=12$nm (blue-square) down to the $D=3$nm (red-dot) NMOS [100] NW. The subband edges of each NW are shifted to the same reference $E=0$eV for comparison purposes. The arrows show the direction of diameter decrease. There are two counteracting mechanisms that affect the carrier velocity in these NWs as the diameter decreases: i) The Γ mass increases, a result of non-parabolicity in the dispersion of the Si bulk bandstructure. From the bulk value of $m^*=0.19m_0$, it increases to $0.27m_0$ at $D=3$nm [13]. ii) The off-Γ valleys with heavier transport mass ($m^*=0.89m_0$), but light quantization mass, shift higher in energy. The second mechanism is slightly stronger, and the combined effect is that the velocities are slightly higher for NWs of smaller diameters. As the gate bias increases, however, electrostatic confinement also increases the valley separation of the larger diameter NWs. No clear trend in the velocities at all biases can, therefore, be identified as earlier described in Fig. 3a (right panel).

Figure 4b shows the first occupied subband of the [110] NWs of all diameters. As the diameter reduces: i) The Γ mass slightly reduces (a result of the anisotropic dispersion of the Si bulk bandstructure [13]), and ii) the heavier transport mass off-Γ valleys shift higher in energy. Both effects tend to increase the carrier velocities. A clear trend in velocity reduction as the diameter reduces is therefore observed, as shown in Fig. 3b (right panel).

In the case of the [111] NMOS NWs in Fig. 4c, the mass of the first conduction subband slightly increases as the diameter reduces (the curvature reduces). This increase is only marginal, and does not lead to any observable variations in the carrier velocities as shown in Fig. 3c - right panel.

Figure 5 shows the same quantities as in Fig. 4, but for the PMOS NWs. Figure 5a shows the first *two* occupied subbands for the [100] PMOS NWs as the diameter reduces from $D=12$nm (blue-square) to $D=3$nm (red-dot). The subbands indicate no clear trend in their curvature as a function of diameter. Instead, an oscillatory behavior is observed [14], with several band-crossings between bands from wires of different diameters. (Showing the higher *two* bands in this case indicates the oscillations more clearly). This reflects in the velocities of Fig. 3a (left panel), for which no significant variation exists, and as the gate bias increases, the magnitude of the velocities of wires with different diameters is also observed to interchange. The oscillatory behavior of the subbands keeps the carrier velocities low. The low bias velocity of [100] PMOS NWs is $v_{inj} \sim 0.5 \times 10^5$ m/s, whereas in all other NW categories the velocities are almost 2 X higher.

The variation pattern in the subband envelopes of the [110] and [111] PMOS NWs in Fig. 5b, 5c is clearer. Here the subbands undergo a large transformation as the diameter decreases, acquiring a larger curvature and lighter effective masses, and thus significantly higher carrier velocities. This explains the velocity trend in Fig. 3b, 3c (left panels). The subband shape behavior, and its large change under cross section reduction is described in detail in Ref. [14], and is a result of the anisotropy of the heavy-hole subband of the valence band shown in the inset of Fig. 5b. There, the 45º lines drawn show the relevant energy lines that form the subbands for NWs with the (1-10) surface quantized, as is the case for the [110] (and [111]) oriented NWs. As the diameter reduces, subbands further away from the center of the Brillouin zone are utilized, which have large curvatures and lower effective masses. The arrow along the 45º lines shows the direction of decreased wire cross section, corresponding to larger k-value quantization. The subband trend in Fig. 5b has its origin in this anisotropic energy surface. Of course, real NW quantization involves many more interactions, but the basic trend of the heavy-hole band is transferred to the NW subbands. A similar effect is responsible for the subband trend of the [111] NWs shown in Fig. 5c.

## V. Results for Rectangular NWs

After investigating the carrier velocity and current variations of cylindrical NWs, we extend our analysis to rectangular NWs with widths/heights varying from 3nm to 12nm (all aspect ratios), for the three orientations under consideration. Figures 6 and 7 show the results for NMOS and PMOS NWs, respectively. Due to the large volume of the data for NWs with various aspect ratios and gate biases, only the velocity (first row) and current (second row) results at on-state ($V_G=V_D=1V$ for NMOS, and $V_G=V_D=-1V$ for PMOS) are presented. The lower left corners of the sub-figures in Fig. 6 and 7 show the velocity/current of the 3nm x 3nm NWs, whereas the upper right corners show the velocity/current of the 12nm x 12nm wires. Other than the width/height of the NWs no other parameter is changed in the simulations.

Figures 6a, 6b and 6c (first row of Fig. 6) present the velocity results for the [100], [110] and [111] oriented NMOS NWs, respectively. In all cases, cross section reduction results in higher velocities (higher velocities in the lower/left than the upper/right part of the figures). In the [100] NW case in Fig. 6a, following the cylindrical case arguments, the off-Γ valley is pushed higher in energy and the overall velocity is higher. The velocity variation in the entire figure is of the order of ~8%, ranging from $1.2 \times 10^5$ m/s to $1.3 \times 10^5$ m/s.

In the case of the [110] NMOS NW in Fig. 6b, the width and the height surfaces are in the [1-10] and [001] directions, respectively, as shown in the third row of Fig. 6. The nature of valley quantization is different for each surface. In the height in the [001] quantization direction, the Γ valleys have *light transport* mass, but *heavy quantization* mass. The off-Γ valleys have the reverse, *heavy transport* mass, but *light quantization* mass. Reducing the height lifts the lightly quantized off-Γ valleys, just as in the [100] NW cases above, and increases the velocities (i.e the velocities increase as one moves from top to bottom in Fig. 6b). On the other hand, in the [1-10] width direction, the off-Γ valleys are more heavily quantized than the Γ valleys. Variations in the width do not shift their energy minima strongly, and the carrier velocities are therefore almost constant along that direction.

An extrapolation of our results, indicates that NMOS (001)/[110] channels (extension beyond the lower-right corner of Fig. 6b), have higher velocities than NMOS (1-10)/[110] channels (extension beyond the upper-left corner of Fig. 6b). Experimental data on the channel mobility vs. transport and quantization orientation in Si MOSFET channels [43, 44] show that the mobility is also higher for NMOS (001)/[110] rather than (1-10)/[110] channels. Although the mobility depends also on the scattering process and not only on bandstructure, the velocity results point towards a possible explanation of the experimental behavior. Furthermore, for long channel NWs devices with finite width, the fact that the velocity is not sensitive to variations in the [1-10] direction, points towards utilizing this direction as the one for which the line etch control is minimal in the fabrication process, so that the performance variation due to size variations is reduced.

Comparing the magnitude of the carrier velocities, it is very similar in the [100] and [110] oriented NWs in Fig. 6a and 6b since in both cases it is mostly determined by the Γ valleys. It ranges from 1.1 to $1.3 \times 10^5$ m/s. On the other hand, the velocities of the [111] oriented NWs in Fig. 6c, are determined by tilted Si conduction band ellipsoids of higher effective mass (*m\** ~

0.43$m_0$ in bulk) [13]. They are therefore ~30% lower, ranging only from 0.9 to 1.1 x $10^5$ m/s. Still, however, the same pattern is followed, where size reduction increases carrier velocities.

Figures 6d, 6e and 6f present the ballistic on-current results for the NWs in the three orientations. Since the charge increases linearly with the cross section (by almost four times in Fig. 2b), and the velocity decreases by only ~30%, the on-current increases monotonically following the increase in the cross section. As in the cylindrical NW cases earlier, the on-current is higher for the [100] NWs, closely followed by the [110] NWs, whereas the [111] NWs have ~25% lower on-current. The contour lines in the sub-figures, all tilted at ~45º, indicate the linear increase in $I_D$ with cross section increase, as well as the symmetry between width/height surfaces, even in the velocity asymmetric case of the [110] NWs.

Figures 7a, 7b and 7c present the carrier velocities for the rectangular PMOS NWs. The velocities of the [110] and [111] NWs in Fig. 7b and 7c range from $v_{inj}$ ~ 1 x $10^5$ m/s to 1.5 x $10^5$ m/s, and $v_{inj}$ ~ 1.2 x $10^5$ m/s to 1.8 x $10^5$ m/s respectively, a variation of ~50%. For the [100] NWs in Fig. 7a, the velocities range from $v_{inj}$ ~ 0.5 x $10^5$ m/s to 0.8 x $10^5$ m/s and are much lower compared to all other NW cases of either carrier type. Similar to the NMOS case, in general, cross section size reduction increases the carrier velocities. This is more evident in the [100] and [111] NW orientation cases of Fig. 7a and Fig. 7c, respectively. Figure 7b, on the other hand, shows a strong surface anisotropic behavior for the [110] PMOS NWs. Scaling of either NW side (width in [1-10] or height in [001]), increases carrier velocities. In the case of scaling the height, however, at ~6nm the velocity gets an upward jump before it starts to decrease again. This is attributed to the anisotropic quantization mass in the two surfaces and detailed explanations are provided in Ref. [15].

It is also worth mentioning here that in the case of the PMOS [110] channels, experimental data [43, 44] show that long channel (1-10)/[110] MOSFET channels have higher mobility than the (001)/[110] ones, the opposite of what is observed in the NMOS [110] channel. The different carrier velocities in the two PMOS channels could point to the reasons that might be responsible for this. Scaling of the [1-10] width quantizes the k-space along the light branch of the anisotropic heavy-hole valley (inset of Fig. 5b). Scaling the [001] height does not provide this advantage. Although this surface difference is only weakly reflected in the velocities of the narrow NWs of Fig. 7b (left vs. lower parts), simulations using real 2D bandstructures could potentially demonstrate the difference in velocities between the two surfaces. In general, however, physical scaling of the channel in directions that utilize the larger curvature regions of the bulk bandstructure is beneficial to the carrier velocity and device performance (as the scaling of the [1-10] width in the PMOS (1-10)/[110] case).

Figures 7d, 7e and 7f show the PMOS NWs ballistic on-current results for the three orientations. In the [100] case in Fig. 7d and [111] case in Fig. 7f, the on-current increases monotonically as the cross section increases, similar to the NMOS cases. In the case of the [110] NWs in Fig. 7e, the on-current has a more complex behavior, following the complex behavior of the velocity in Fig. 7b. Here, the regions of 3nm-5nm and 7nm-12nm of height, and for any width are design regions for low on-current variations to large cross section variations. Around a height of ~6nm, however, large variations are observed, and device designs with such height should be avoided. Comparing the magnitude of the on-current in the PMOS NW cases, it is in general higher for the [111] NWs, closely followed by the [110] NWs, whereas the performance of the [100] NWs is almost half compared to the other NWs.

## VI. Design Considerations

Table 1 summarizes the performance comparisons between the NWs of the different orientations for the small ($D$=3nm) and the larger ($D$=12nm) diameters. The indications *"High"*, *"Fair"*, and *"Low"*, refer to the relative performance of the NWs of each row (orientation comparison) and not necessarily on an absolute scale. The numbers within brackets correspond to the performance order (both carrier velocity and current have the same order) of the different orientated NWs within each row. Although this table is constructed according to the cylindrical NW results, the same conclusions follow in the cases of the rectangular devices.

In the case of NMOS NWs, in Table 1a, both the [100] and [110] orientations have high performance, with the [100] orientation having an advantage at larger diameters and the [110] at smaller diameters. The [111] NWs have lower performance at smaller diameters and fair at larger diameters compared to the two other orientations. The PMOS performance comparison is shown in Table 1b. In this case, the [111] orientation is the most advantageous in all diameter ranges, closely followed by the [110] orientation, whereas the [100] orientation performs purely for all NW diameters. PMOS [111] NWs perform higher than all other NWs (PMOS or NMOS), and are the optimal solution for applications that require high performance individual NWs. Since the NMOS [111] and especially the PMOS [100] NWs perform purely, for CMOS applications that both NMOS and PMOS high performance is required, the [110] orientation seems to be the optimal solution.

Up to this point our analysis considered infinitely long, undistorted NWs with perfect surfaces, assuming perfect manufacturability. In reality, structure imperfections exist in devices and affect the performance [45, 46, 47, 48, 49]. Controlling the line edge roughness in nanofabrication processes imposes challenges, and the lack of it leads to device-to-device performance variations. The width is usually less well controlled since it is formed by etching, whereas the height is controlled by growth and can be more precise. For the high performance rectangular PMOS [110] NWs build on (001) substrates, the on-current is not significantly sensitive to the width (Fig. 7e). It is also not significantly sensitive to the height, except at around 6nm of height, a design region that should be avoided. In the NMOS [110] case the (001)/[110] configuration is also beneficial since the velocity is almost constant in the [1-10] width direction. Although the ballistic on-current in Fig. 6e is symmetric with respect to the width/height, devices are not 100% ballistic, and therefore it is still beneficial to have the direction of least control aligned with the direction of velocity invariance.

As a design strategy, therefore, our results demonstrate that out of all NWs examined, the [110] oriented NWs are advantageous for CMOS technology applications in either design case: i) if the design goal is driven by the highest performance assuming perfect manufacturing abilities, or ii) if the design is driven by low device-to-device performance fluctuations. Design regions with velocity and on-current insensitivity to geometry can provide low device-to-device variations strategies for both, long channel devices, and short channel close-to-ballistic devices. In the cases of the [110] channel orientations built on (001) substrates, design regions can be identified for either case, while still keeping the performance high.

An important source of performance variation and degradation is surface roughness scattering (SRS). To quantify the previous results, we examine the effect of SRS on the velocity and current variations for the case of the NMOS [110] $D$=3nm cylindrical NW. For this purpose, real-space 3D, full-band, quantum ballistic simulations are performed using the OMEN code [17, 35, 36], in which the roughness is treated in a realistic way by adding/subtracting atoms on the surface of the NW [45]. A ~15% degradation, as well as an additional ~30% variation in the velocities is computed by using a sample size of 50 rough NWs. A 10% variation in the on-current is also observed, in agreement with other works [46, 48, 49]. This is an indication of an additional variation, on top of the variation expected due to bandstructure as a result of diameter change. A proper and more elaborate investigation the effect of surface roughness scattering which targets all orientations and a large range of diameter sizes will be presented elsewhere. These numbers, however, further stress the need of device and circuit designs tolerant to performance variations originating from structure variations and non-idealities. Bandstructure features can provide a mechanism to partially help on this.

## VII. Conclusion

An atomistic tight-binding approach and a semi-classical ballistic model are used to calculate the bandstructure velocity and ballistic current of nanowire devices, self-consistently with the electrostatic potential. NMOS and PMOS NWs of diameters from $D$=12nm down to $D$=3nm in [100], [110], and [111] orientations, of cylindrical and rectangular cross sectional shapes are considered. The carrier velocities are strong function of orientation, band type, diameter, and bias. Cross section scaling, in general increases the carrier velocities either by raising the energy minima of the heavier transport mass valleys, or by significantly changing the subbands' curvature. PMOS [110] and [111] NWs have the largest velocity sensitivity to diameter, in which cases the velocity doubles as the diameter scales from $D$=12nm to $D$=3nm. The carrier velocity of NMOS [110] NWs is also sensitive to the diameter, but at a smaller degree. On the other hand, the velocities of NMOS [100] and [111], and PMOS [100] NWs are insensitive to the diameter. Gate bias also tends to increase carrier velocities by ~50%, as higher energy states are occupied at inversion conditions. Trends in carrier velocity with diameter do not completely reflect to the terminal current characteristics because the carrier density is also subject to the change of the cross section geometry. The ballistic on-current shows less sensitivity with cross section fluctuations in the cases of [110] PMOS and [111] PMOS NWs, whereas in the rest of the cases it varies linearly with cross section. The PMOS [111] NWs are the ones with the highest performance from all NW categories, whereas the PMOS [100] the ones with the lowest performance. The [110] oriented NWs, on the other hand, are the ones with both, high NMOS and PMOS carrier velocities and on-current, and therefore more suitable for CMOS applications. In either case, if the design goal is driven by the highest performance assuming perfect manufacturability, or if the design is driven by low device-to-device performance fluctuations, this study suggests the (001)/[110] oriented NWs for both NMOS and PMOS NW devices for either long channel, or short channel ballistic devices.

## Acknowledgements:

This work has been partially supported by funds from the Austrian Science Fund, FWF, contract I79-N16. S.G. Kim was supported by Materials Structures and Devices Focus Center MSD/MARCO. Dr. Mathieu Luisier is acknowledged for providing the quantum transport code for the surface roughness calculations and for various discussions. Computational resources of the Network for Computational Nanotechnology (NCN) operated by nanoHUB.org are acknowledged. The simulations in this work can be duplicated with Bandstructure Lab on nanoHUB.org [50].

Figure 1: Dispersions of [110] NMOS and PMOS NWs of $D$=6nm

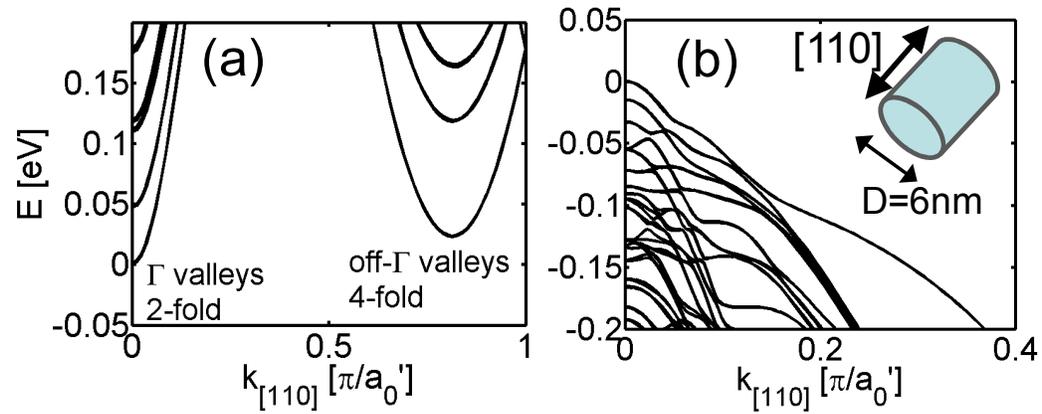

Figure 1 Caption:

Equilibrium dispersion relations for the $D$=6nm [110] NW (positive k-states). (a) Electrons. (b) Holes. The edges of the first band are shifted to $E$=0eV. $a_0$' is the unit cell length.

Figure 2: velocity/charge/$I_{ON}$ variations in cylindrical NMOS/PMOS NWs at $|V_G|=1V$ vs. diameter.

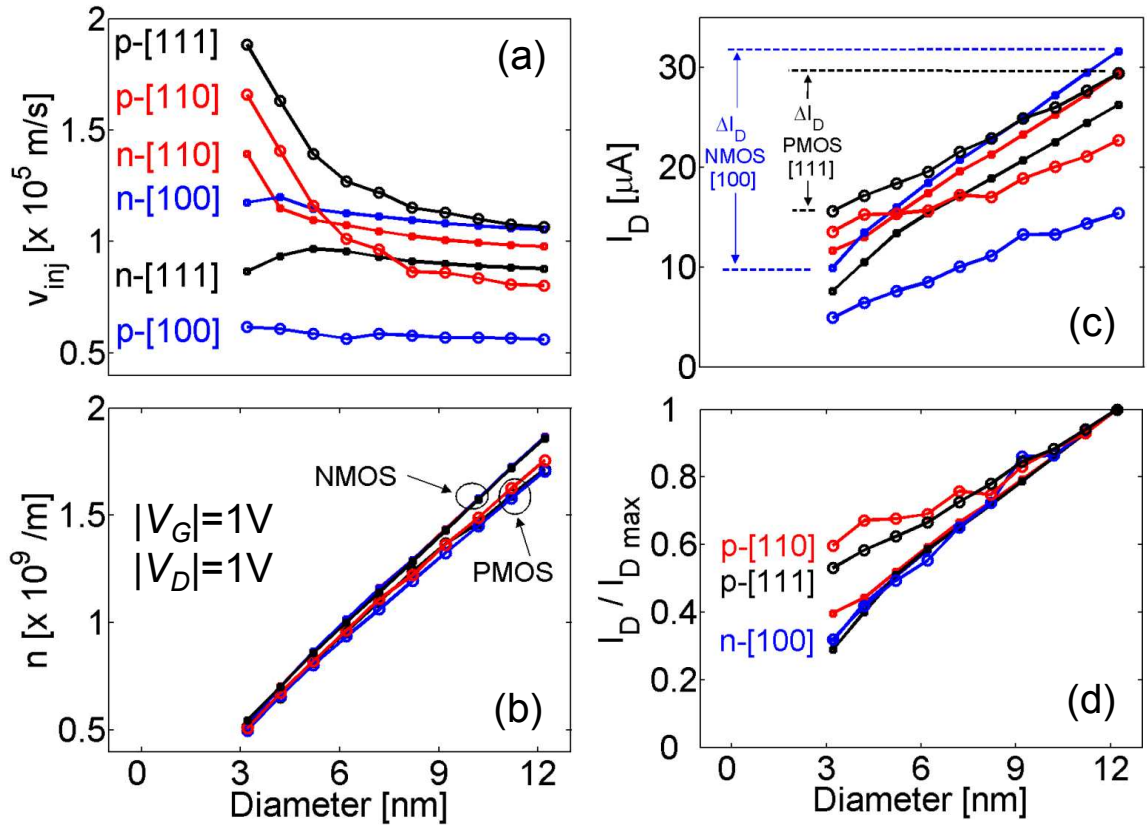

Figure 2 Caption:

Results for cylindrical NMOS and PMOS NWs at $|V_G|=1V$ and $|V_D|=1V$ vs. diameter. (a) Carrier injection velocity. (b) Charge density. (c) On-current. (d) On-current normalized to its maximum value (the $D$=12nm value).

Figure 3: velocity/charge/current variations in cylindrical NMOS/PMOS for different diameters vs. bias

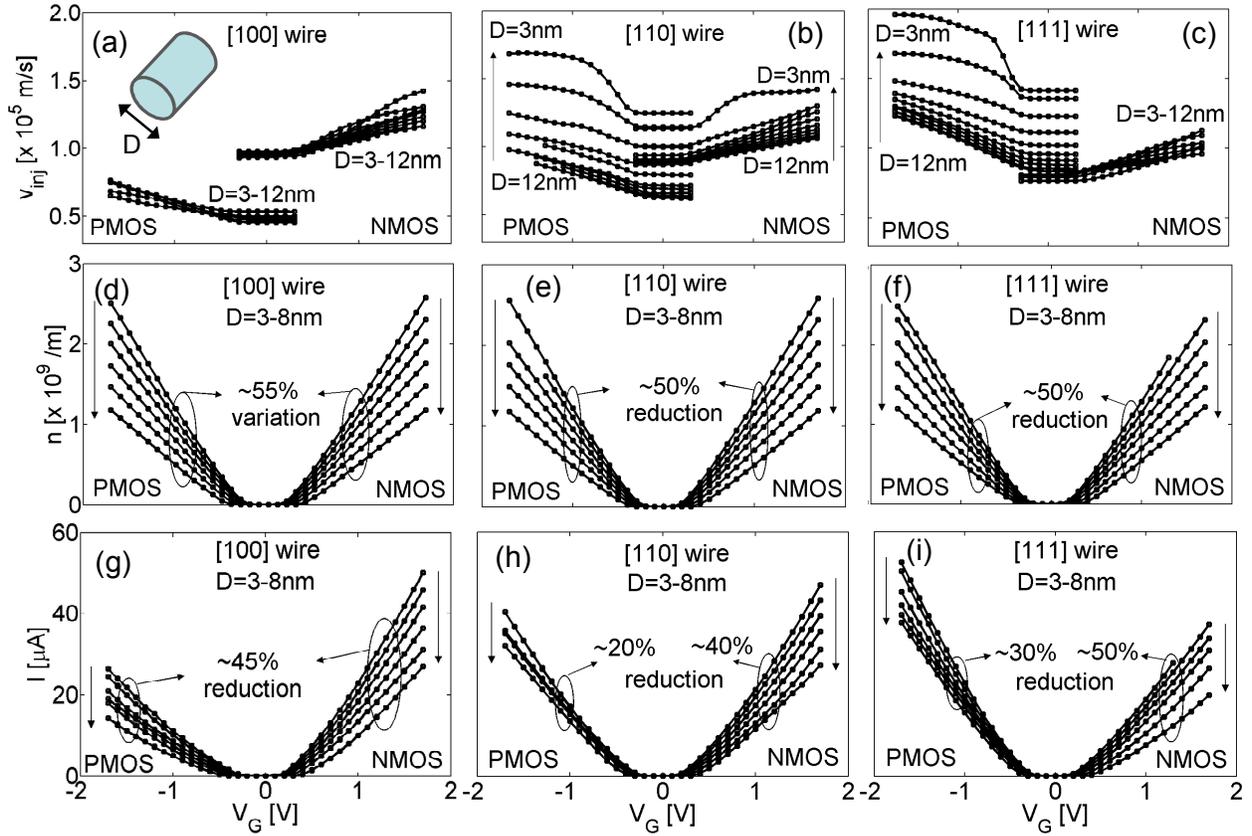

Figure 3 Caption:

Results for cylindrical NWs of diameters varying from $D$=12nm down to 3nm, PMOS and NMOS vs. $V_G$ for $|V_D|$=1V. PMOS NW results are shown for negative $V_G$, NMOS results for positive $V_G$. The arrows indicate the direction of diameter decrease in the cases for which a uni-directional trend is observed. (a-c) Carrier velocities. (d-f) Charge density (for $D$=8nm down to $D$=3nm). (g-i) Ballistic current (for $D$=8nm down to $D$=3nm). Left column - [100] oriented wires. Middle column - [110] oriented wires. Right column - [111] oriented wires. Inset of (a): The cylindrical NW cross section.

Figure 4: The first subband (subband envelope) of NMOS NWs as the diameter changes

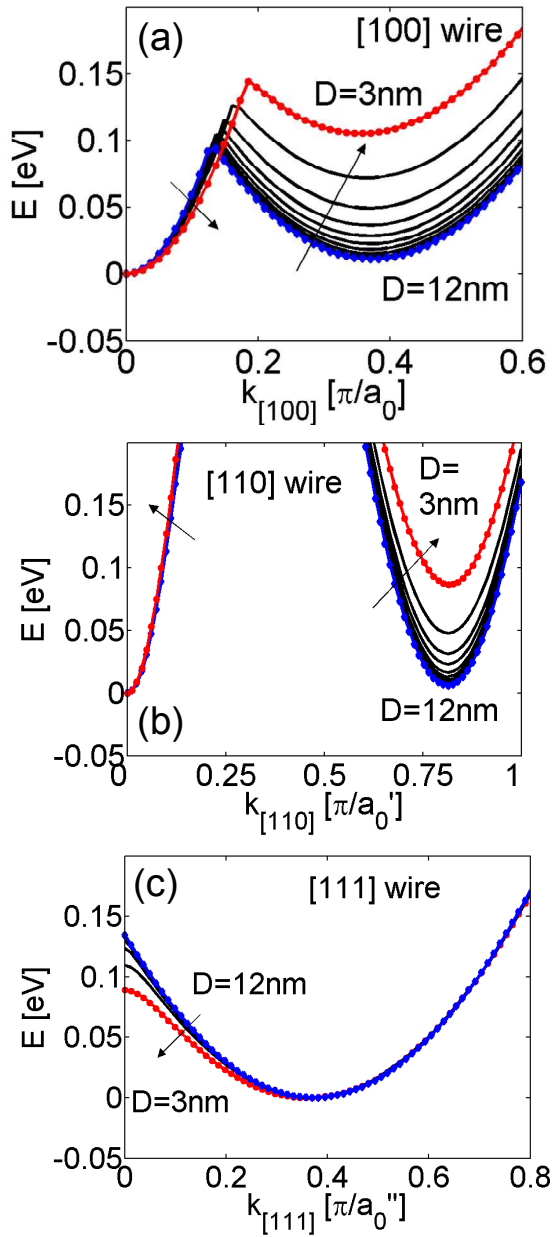

Figure 4 Caption:

The first subband (subband envelope) of NMOS NWs as the diameter scales from $D$=12nm to 3nm. The arrows indicate the direction of diameter decrease. (a) [100] oriented wires. (b) [110] oriented wires. (c) [111] oriented wires. The minima of all bands are shifted to $E$=0 eV.

Figure 5: The first subband (subband envelope) of PMOS NWs as the diameter changes

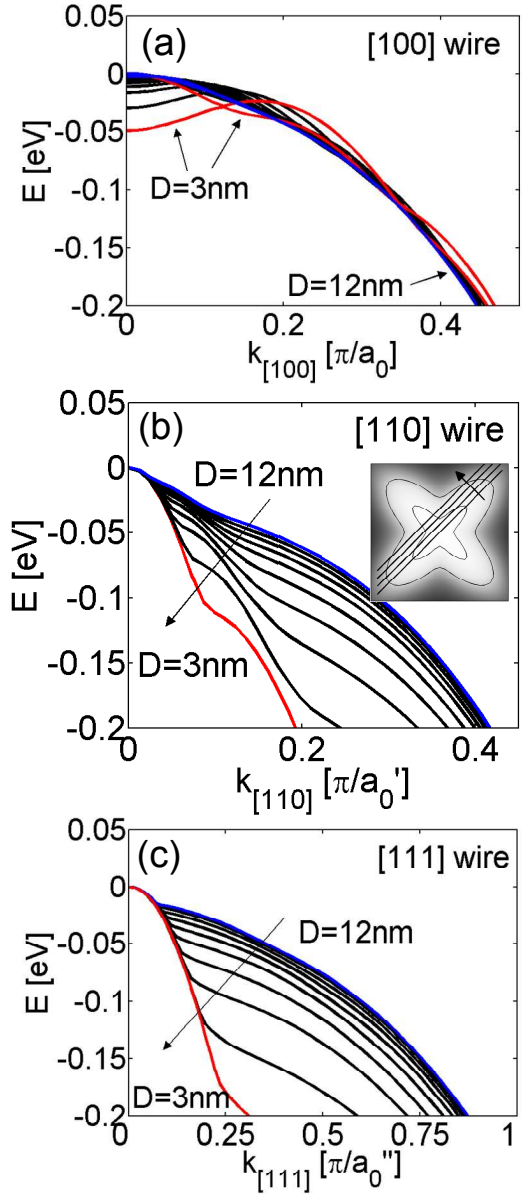

Figure 5 Caption:

The first subband (subband envelope) of PMOS NWs as the diameter scales from $D$=12nm to 3nm. The arrows indicate the direction of diameter decrease. (a) [100] oriented wires (the first two subbands are shown). (b) [110] oriented wires. (c) [111] oriented wires. The minima of all bands are shifted to $E$=0 eV. Inset of (b): The heavy-hole subband of the Si bandstructure with (1-10) surface quantization subbands indicated.

Figure 6: The average velocity and current of rectangular NMOS NWs at ON-state

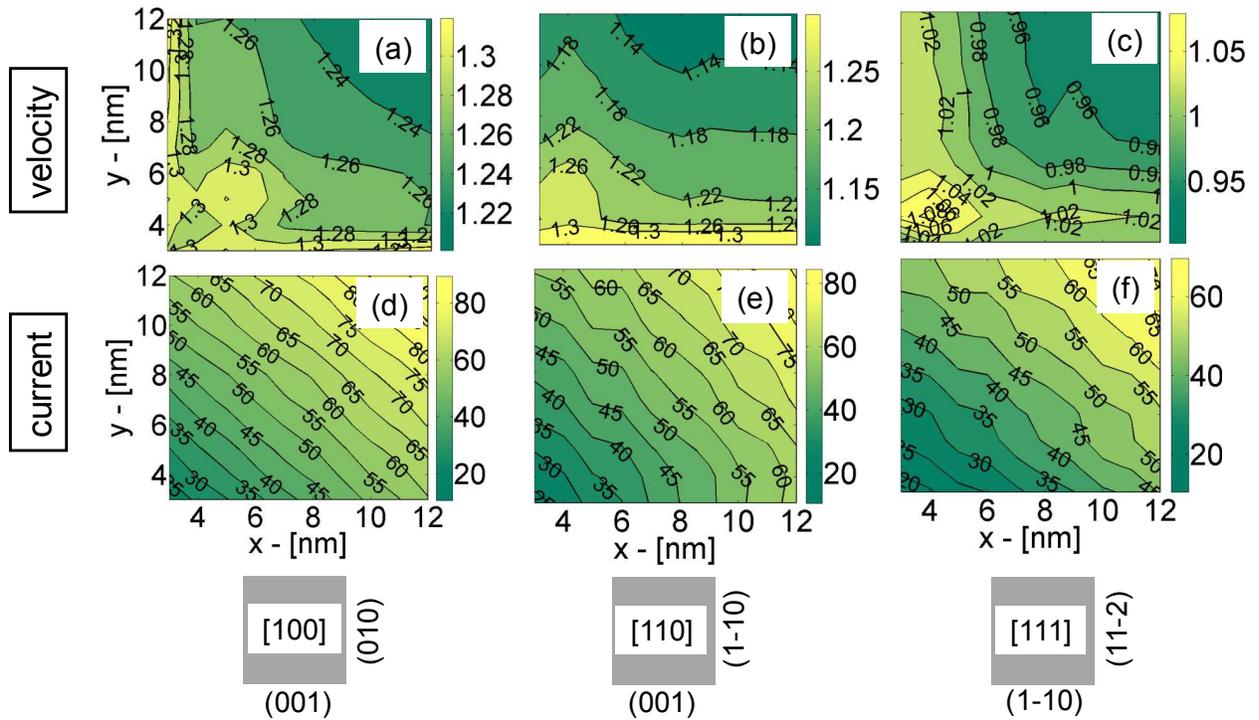

Figure 6 Caption:

2D surfaces of transport quantities for rectangular NMOS NWs as the width (*W*) / height (*H*) directions vary from *W*=3nm to *W*=12nm, and *H*=3nm to *H*=12nm. (a-c) The average carrier velocity in $10^5$ m/s for [100], [110] and [111] directed wires respectively. (d-f) The current density in µA for [100], [110] and [111] directed wires respectively. The lower row indicates the NW directions and surface orientations.

Figure 7: The average velocity and current of rectangular PMOS NWs at ON-state

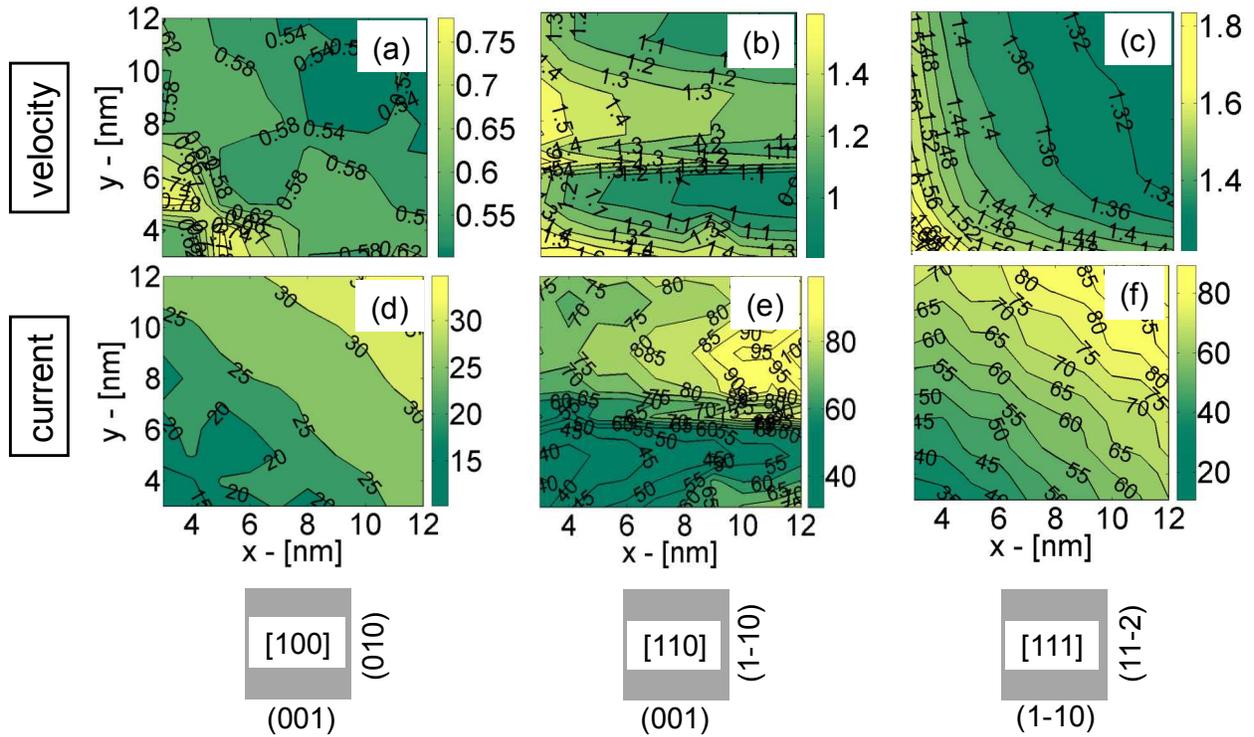

Figure 7 Caption:

2D surfaces of transport quantities for rectangular PMOS NWs as the width ($W$) / height ($H$) directions vary from $W$=3nm to $W$=12nm, and $H$=3nm to $H$=12nm. (a-c) The average carrier velocity in $10^5$ m/s for [100], [110] and [111] directed wires respectively. (d-f) The current density in µA for [100], [110] and [111] directed wires respectively. The lower row indicates the NW directions and surface orientations.

Table 1: NMOS and PMOS comparisons

| **NMOS** performance | [100] | [110] | [111] |
|---|---|---|---|
| Small $D$ (3nm) | High (2) | High (1) | Low (3) |
| Large $D$ (12nm) | High (1) | High (2) | Fair (3) |

(a)

| **PMOS** performance | [100] | [110] | [111] |
|---|---|---|---|
| Small $D$ (3nm) | Low (3) | High (2) | High (1) |
| Large $D$ (12nm) | Low (3) | Fair (2) | High (1) |

(b)

Table 1 Caption:

Relative performance comparison for the NWs of the different orientations and diameters: (a) NMOS. (b) PMOS.